\pdfoutput=1

\documentclass[11pt]{article}

\usepackage[]{ACL2023}

\usepackage{times}
\usepackage{latexsym}
\usepackage{amsfonts}
\usepackage{amsmath}
\usepackage{amssymb}
\usepackage{booktabs}
\usepackage{makecell}
\usepackage[frozencache=true,cachedir=mintedcache]{minted}
\usepackage{float}
\usepackage{mwe}

\newfloat{Code Snippet}{htbp}{loc}
\newfloat{Example}{htbp}{loc}

\usepackage[T1]{fontenc}

\usepackage[utf8]{inputenc}

\usepackage{microtype}
\usepackage{inconsolata}

\usepackage{graphicx}
\usepackage{subcaption}
\usepackage{adjustbox}
\usepackage{array} %

\usepackage{caption}

\usepackage{afterpage} %
\usepackage{floatpag}

\usepackage{xparse}

\title{Tutorial: $\varphi$-Transductions in \textsc{OpenFst} via the Gallic Semiring}

\author{Marco Cognetta \\
  Google \\
  \texttt{cognetta@google.com} \\\And
  Cyril Allauzen \\
  Google \\
  \texttt{allauzen@google.com} \\}

\begin{document}
\maketitle
\begin{abstract}

\textsc{OpenFst}, a popular finite-state transducer library, supports $\varphi$-transitions but, due to an implementation constraint, they cannot be used with transducers in a straightforward way.

In this short tutorial, we describe how one can use other functionality provided by \textsc{OpenFst} (namely, the \textit{Gallic semiring}) to correctly implement $\varphi$-transductions and demonstrate it by implementing the MaxMatch (WordPiece) tokenization algorithm  \cite{devlin-etal-2019-bert,song-etal-2021-fast}. Accompanying self-contained code examples are provided.\footnote{\url{https://www.openfst.org/twiki/pub/Contrib/FstContrib/phi_transduction_tutorial_code.tgz}}

\end{abstract}

\section{Transduction and $\varphi$-Transitions}

Here, we provide a brief example of what we want to accomplish. $\varphi$-transitions are used to encode "fallback" transitions at a given state in a transducer. That is, when searching for a matching transition at a state (for example, during finite-state transducer composition), if no matching arc is found (i.e., there is no arc with a given input label), then an arc with the $\varphi$ label can be traversed. 

\begin{figure*} %
    \centering %

    \begin{subfigure}[t]{0.3\textwidth} %
        \centering
        \includegraphics[width=0.75\textwidth]{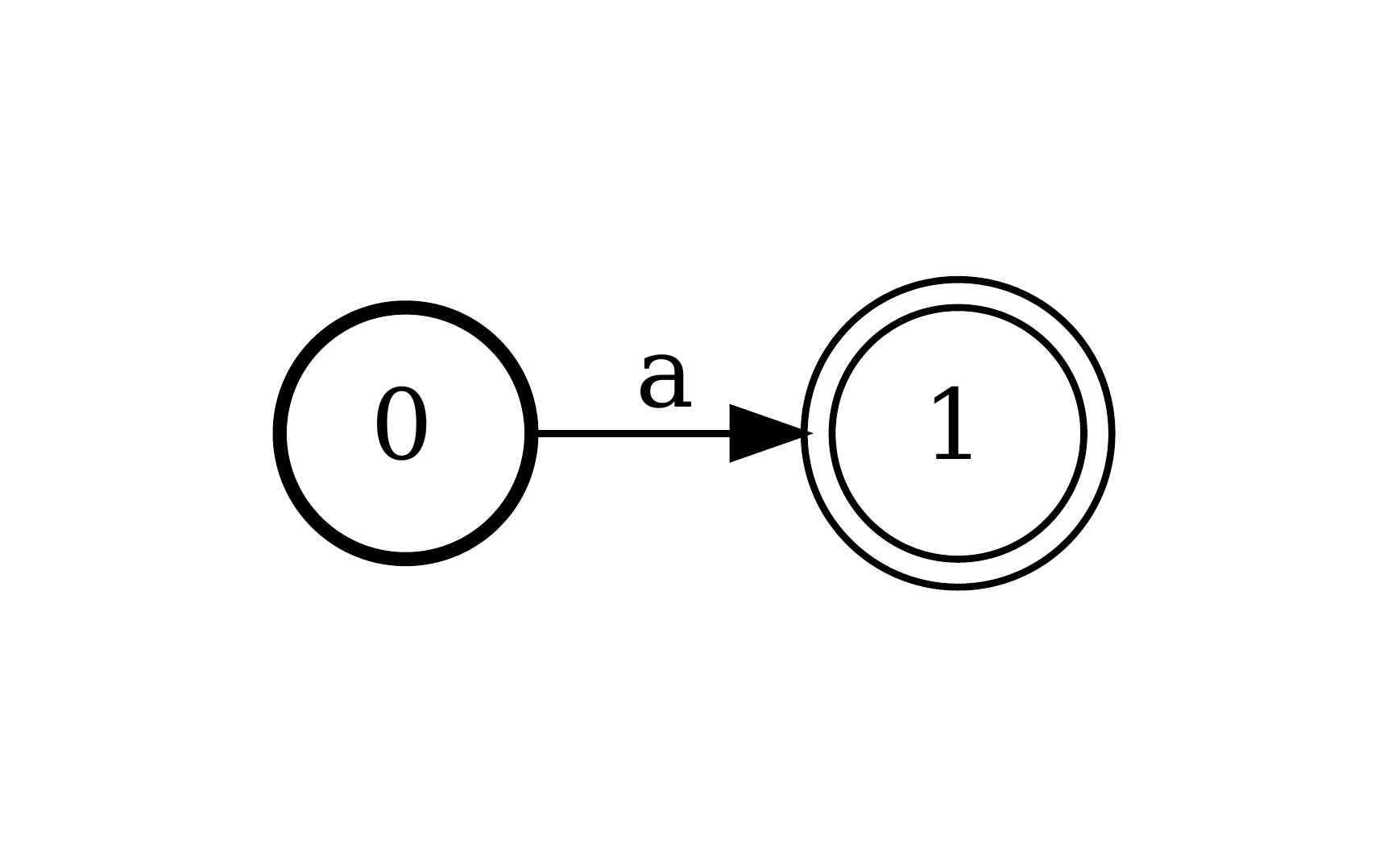} %
        \caption{An automaton encoding \texttt{a}.}
        \label{fig:sub1}
    \end{subfigure}%
    \hfill %
    \begin{subfigure}[t]{0.3\textwidth}
        \centering
        \includegraphics[width=\textwidth]{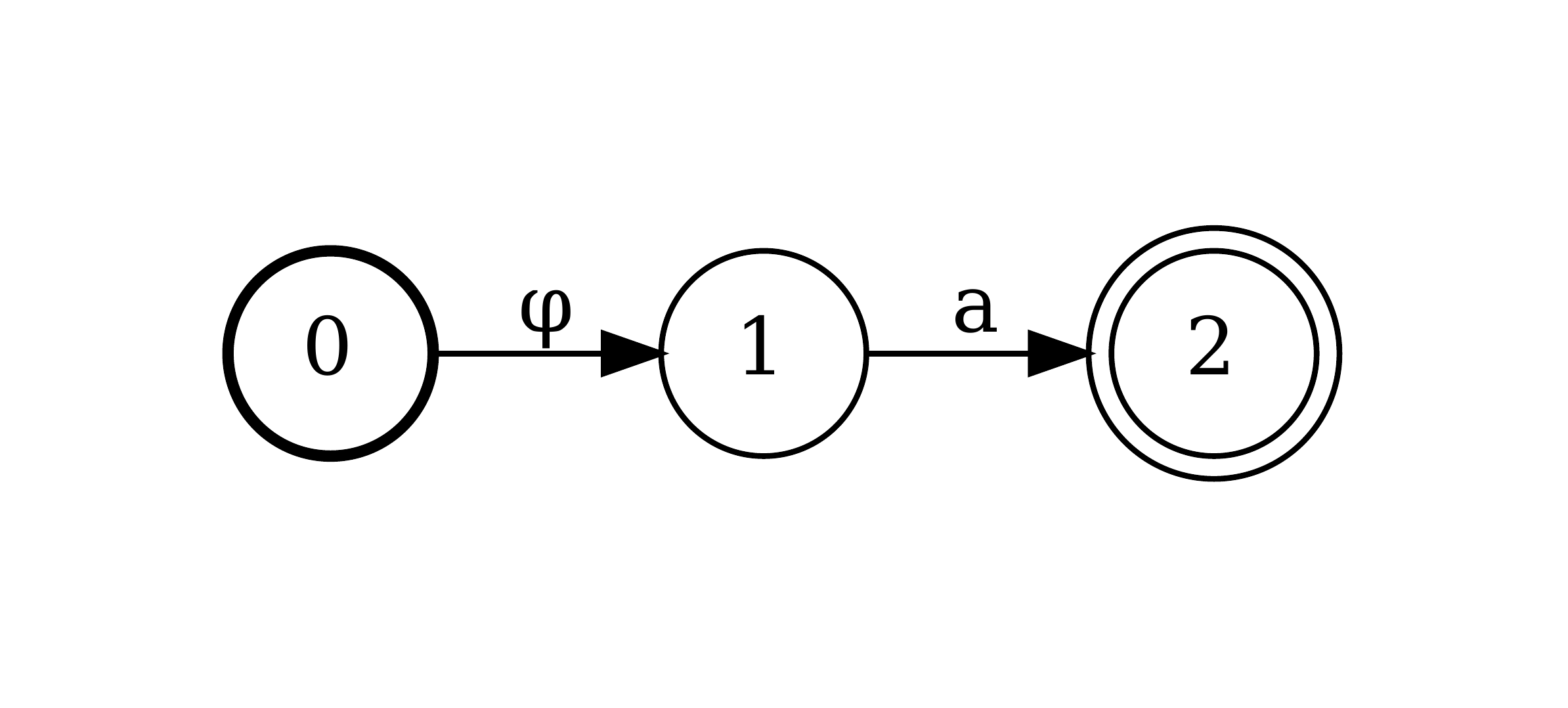} %
        \caption{An automaton that accepts $\texttt{a}$.}
        \label{fig:sub2}
    \end{subfigure}%
    \hfill
    \begin{subfigure}[t]{0.3\textwidth}
        \centering
        \includegraphics[width=0.75\textwidth]{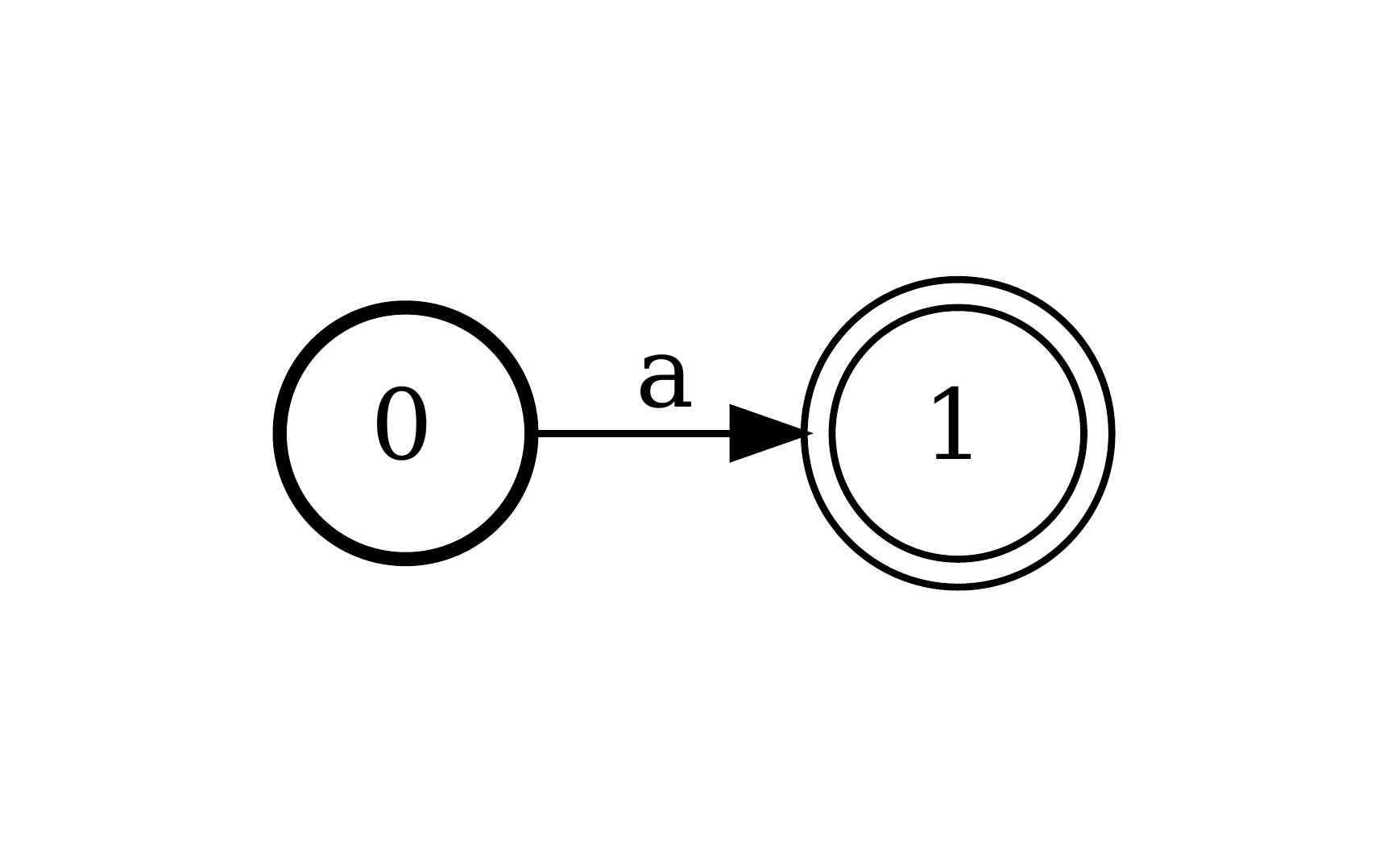} %
        \caption{The composition of (a) and (b).}
        \label{fig:sub3}
    \end{subfigure}

    \vspace{\baselineskip} %

    \begin{subfigure}[t]{0.3\textwidth}
        \centering
        \includegraphics[width=0.75\textwidth]{simple_example/input_pattern_automaton.pdf} %
        \caption{An automaton encoding \texttt{a}.}
        \label{fig:sub4}
    \end{subfigure}%
    \hfill
    \begin{subfigure}[t]{0.33\textwidth}
        \centering
        \includegraphics[width=\textwidth]{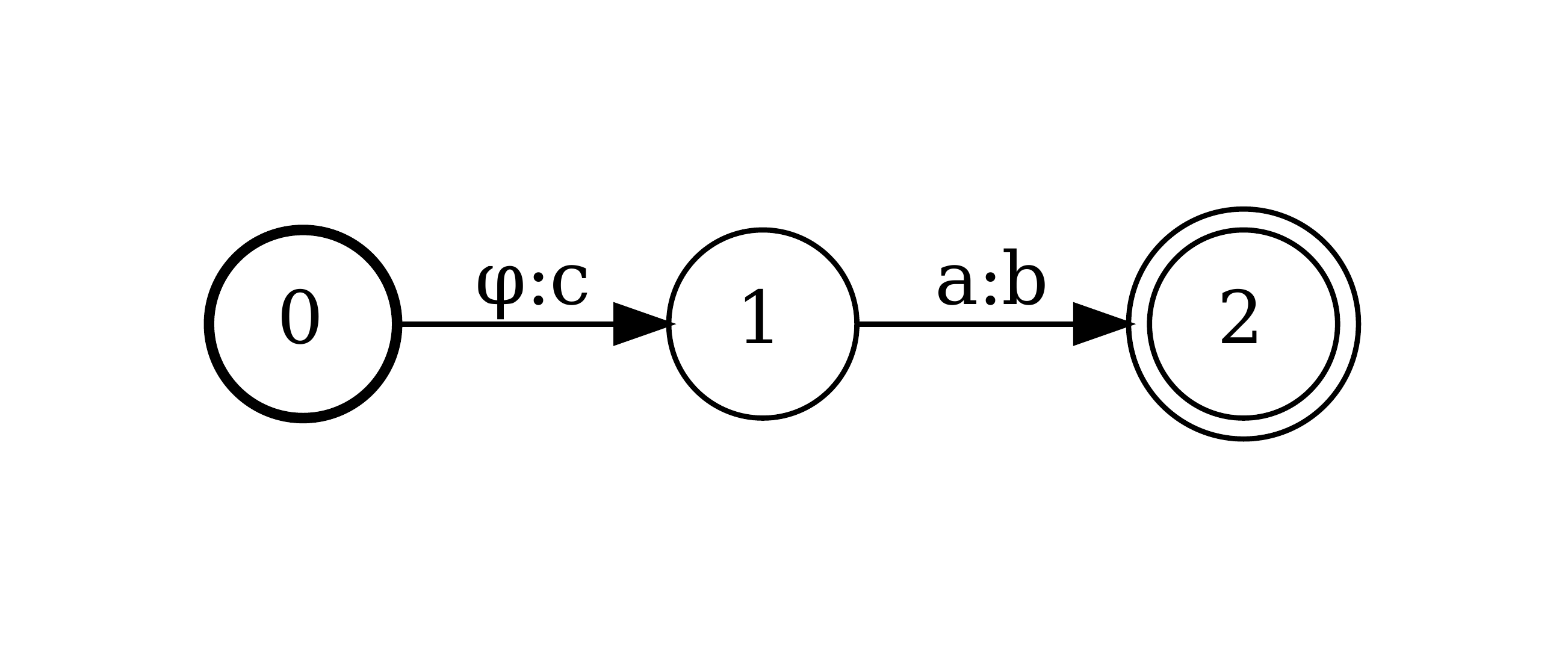} %
        \caption{A transducer for $\texttt{[a]} \rightarrow \texttt{[c][b]}$.}
        \label{fig:sub5}
    \end{subfigure}%
    \hfill
    \begin{subfigure}[t]{0.3\textwidth}
        \centering
        \includegraphics[width=\textwidth]{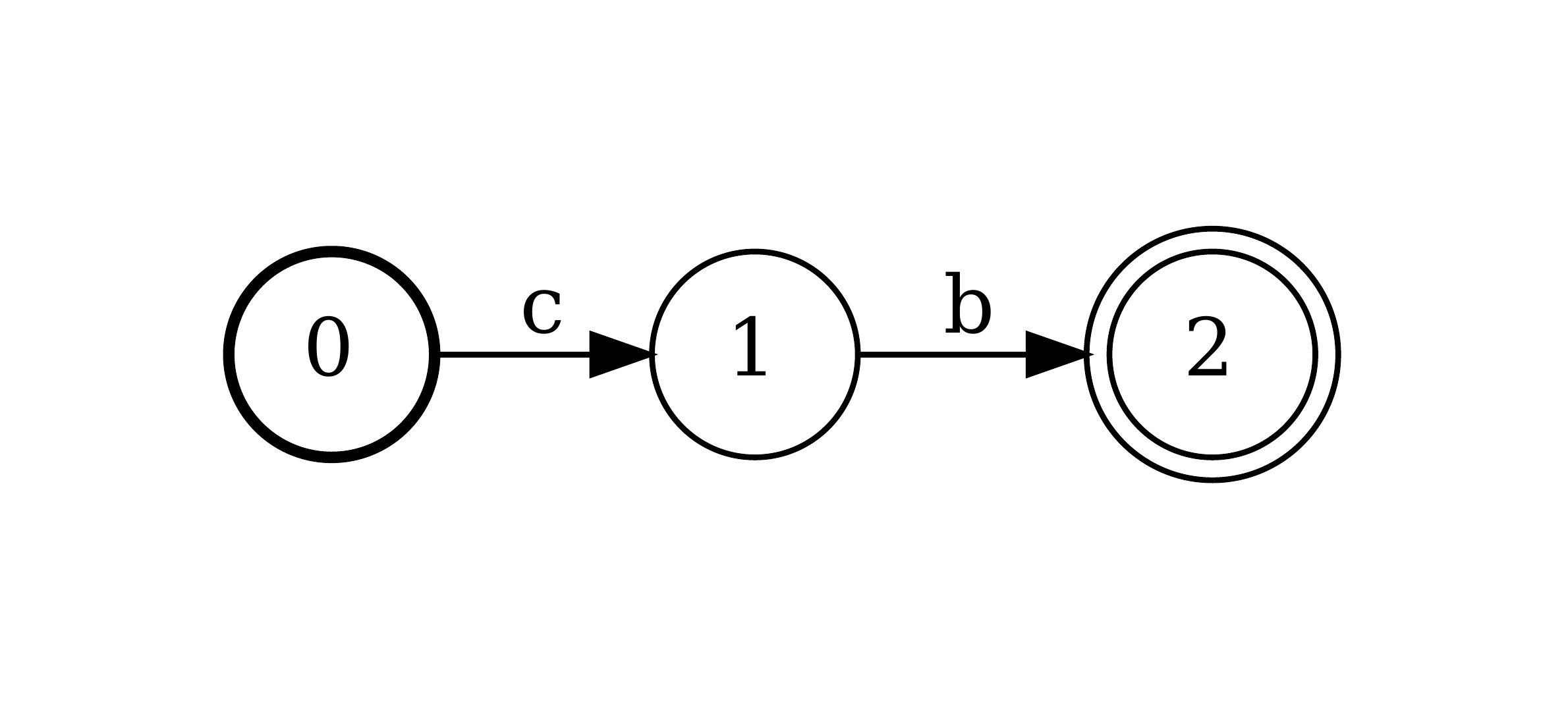} %
        \caption{The composition of (d) and (e), after projecting to the output labels.}
        \label{fig:sub6}
    \end{subfigure}

    \caption{Example inputs and outputs for composition with $\varphi$-transitions. Note that the result of composing (d) and (e) using the na\"ive composition method in Code Snippet \ref{code:incorrect} incorrectly produces the empty language (but the automaton case, implemented in Code Snippet \ref{code:phi_automata}, works as expected).}
    \label{fig:simple_example}
\end{figure*}

Consider the automata and transducers in \ref{fig:simple_example}, each over the alphabet $\Sigma = \{\texttt{a}, \texttt{b}, \texttt{c}\}$. In Figures \ref{fig:sub1}-\ref{fig:sub3}, we have two automata that get composed together. The left automaton accepts only the language $\texttt{a}$, while the right automaton accepts the language $\Sigma$ but implemented using just a $\varphi$-transition, which can be traversed by anything in $\Sigma$, since they don't have explicitly-defined transitions at the start state. Thus, the expected output of the composition is the language $\texttt{a}$.

On the other hand, in Figures \ref{fig:sub4}-\ref{fig:sub6}, we again have an automaton that encodes the language $\texttt{a}$ which we will compose with a transducer that encodes the transduction $\texttt{[a]} \rightarrow \texttt{[c][b]}$. This is encoded by the $\varphi\texttt{:c}$ transition from the start state, which can be traversed if no matching symbol is found (and outputs $\texttt{[c]}$). Then, at state $1$, we find a matching $\texttt{a:b}$ transition and traverse that.

However, while the na\"ive implementation of \textit{automata} composition with $\varphi$-transitions works in \textsc{OpenFst}, the \textit{transducer} case fails and outputs an incorrect language. We give an example of code to reproduce these examples in Code Snippet \ref{code:phi_automata} and \ref{code:incorrect}.\footnote{We deliberately do not include it here in the main text so as to not confuse the reader.} A brief explanation of the design choices in \textsc{OpenFst} that cause this error is given in Appendix \ref{apx:why_doesnt_it_work}, but the key reason is that \textsc{OpenFst}'s \texttt{PhiMatcher} implementation discards output labels when traversing $\varphi$-transitions.

\begin{table*}[t]
  \centering
  \small
  \begin{tabular}{l|c} \toprule
    \textit{$\varphi$-transitions} & \makecell{Transitions that can only be taken if there is no \\ matching character at the current state.} \\ \midrule
    \textit{Gallic Semiring} & \makecell{A product semiring over any semiring and \\ the string semiring.} \\ \midrule
    \texttt{ArcMap[Fst]} & \makecell{A tool in \textsc{OpenFst} to transform arcs to a new type \\ via a simple mapping.} \\ \midrule
    \texttt{PhiMatcher} & A tool in \textsc{OpenFst} that helps traverse $\varphi$-transitions. \\ \midrule 
    \texttt{FactorWeight[Fst]} & \makecell{Similar to an arc mapper but transforms a single arc into \\ a sequence of arcs according to some function.} \\ \bottomrule
  \end{tabular}
  \caption{A quick-reference glossary of terms.}
  \label{tab:glossary}
\end{table*}

\section{Transducers and \textsc{OpenFst}}\label{sec:transducers}

We assume that the reader is generally familiar with finite-state transducers and \textsc{OpenFst}, so we do not provide a complete overview here. Instead, we review the key tools that are used in this tutorial and direct the reader to \textsc{OpenFst}'s resources for a more general overview of the library.\footnote{\url{https://www.openfst.org/}} \cite{DBLP:conf/wia/AllauzenRSSM07} 

\subsection{$\varphi$-transitions and Matchers\footnote{\url{https://www.openfst.org/twiki/bin/view/FST/FstAdvancedUsage\#FstMatchers}}}

$\varphi$-transitions are a special \textit{failure} transition. During traversal or composition, an arc with a $\varphi$ label can only be traversed if no matching arc for the current consuming symbol is found at the current state. $\varphi$-transitions are \textit{non-consuming}, similar to $\varepsilon$-transitions, except that $\varepsilon$-transitions do not have any restrictions on traversal. 

A special \texttt{PhiMatcher} class is provided by \textsc{OpenFst} for matching $\varphi$-transitions during traversal or composition.

\subsection{Gallic Semiring\footnote{\url{https://www.openfst.org/twiki/bin/view/FST/FstAdvancedUsage\#Weights}}}

Let $(\mathbb{S}, \oplus_{\mathbb{S}}, \otimes_{\mathbb{S}}, 0_{\mathbb{S}}, 1_{\mathbb{S}})$ and $(\mathbb{T}, \oplus_{\mathbb{T}}, \otimes_{\mathbb{T}}, 0_{\mathbb{T}}, 1_{\mathbb{T}})$ be semirings. Then, a product semiring $\mathbb{P}$ is defined as $(\mathbb{S} \times \mathbb{T}, \oplus_{\mathbb{P}}, \otimes_{\mathbb{P}}, 0_{\mathbb{P}}, 1_{\mathbb{P}})$ where $(a, b) \oplus_{\mathbb{P}} (c, d) = (a \oplus_{\mathbb{S}} c, b \oplus_{\mathbb{T}} d)$ for $(a, b), (c, d) \in \mathbb{S} \times \mathbb{T}$, and likewise for $\otimes_{\mathbb{P}}$. $0_{\mathbb{P}}$ and $1_{\mathbb{P}}$ are defined as $(0_{\mathbb{S}}, 0_{\mathbb{T}})$ and $(1_{\mathbb{S}}, 1_{\mathbb{T}})$, respectively.

The string semiring is a (left\footnote{A \textit{left} semiring only distributes on the left side.}) semiring over an alphabet $\Sigma^*$. Here, $a \oplus b$ is defined as the longest common prefix of $a$ and $b$ and $a \otimes b$ is defined as $ab$ (concatenation). The zero element is an ``infinite'' string such that $a \oplus 0 = 0 \oplus a = a$ (this element exists only to satisfy the constraints of a semiring) and the one element is the empty string.

The \textit{Gallic semiring} is a product semiring where $\mathbb{S}$ is the string semiring and $\mathbb{T}$ is any semiring. The Gallic semiring is implemented in \textsc{OpenFst} via the \texttt{GallicWeight} and \texttt{GallicArc} types.

\begin{figure*}[p!]
        \thisfloatpagestyle{empty}
\vspace*{-1cm}
\captionsetup[subfigure]{justification=centering}
  \begin{tabular}[b]{cc}
    \begin{tabular}[b]{c}
            \begin{subfigure}[b]{0.65\columnwidth} %
                
                \hspace*{-0.8cm}\includegraphics[scale=0.1]{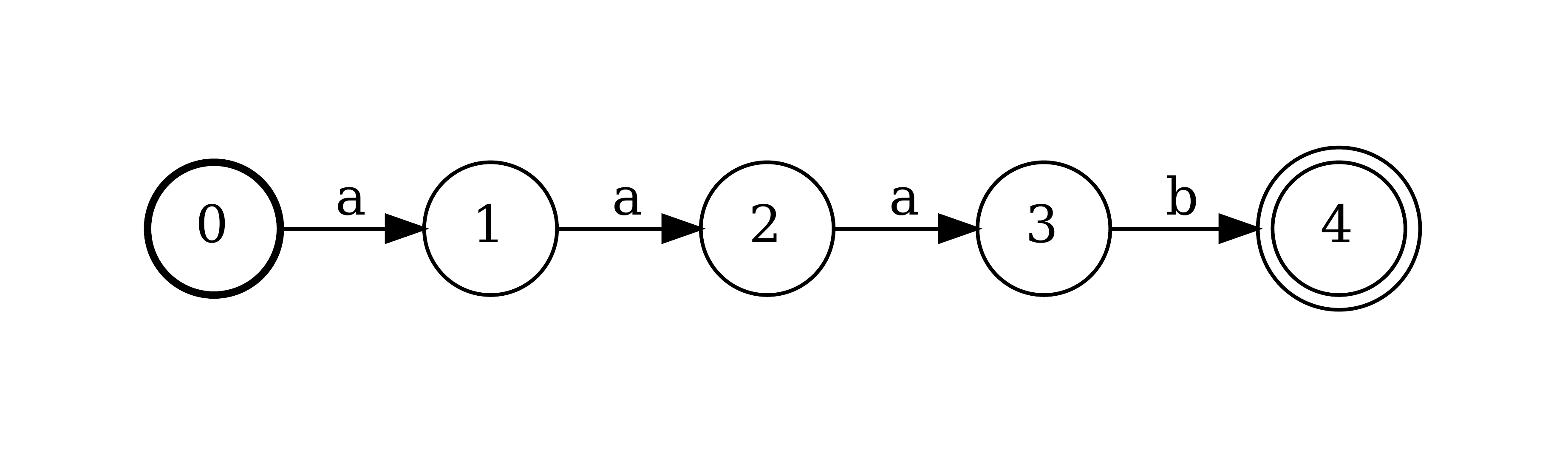} %
                \caption{The input pattern automaton. (\texttt{pattern})}
                \label{fig:A1}
            \end{subfigure}
            \hfill
            \begin{subfigure}[b]{0.41\columnwidth} %
                \centering
                \includegraphics[scale=0.1]{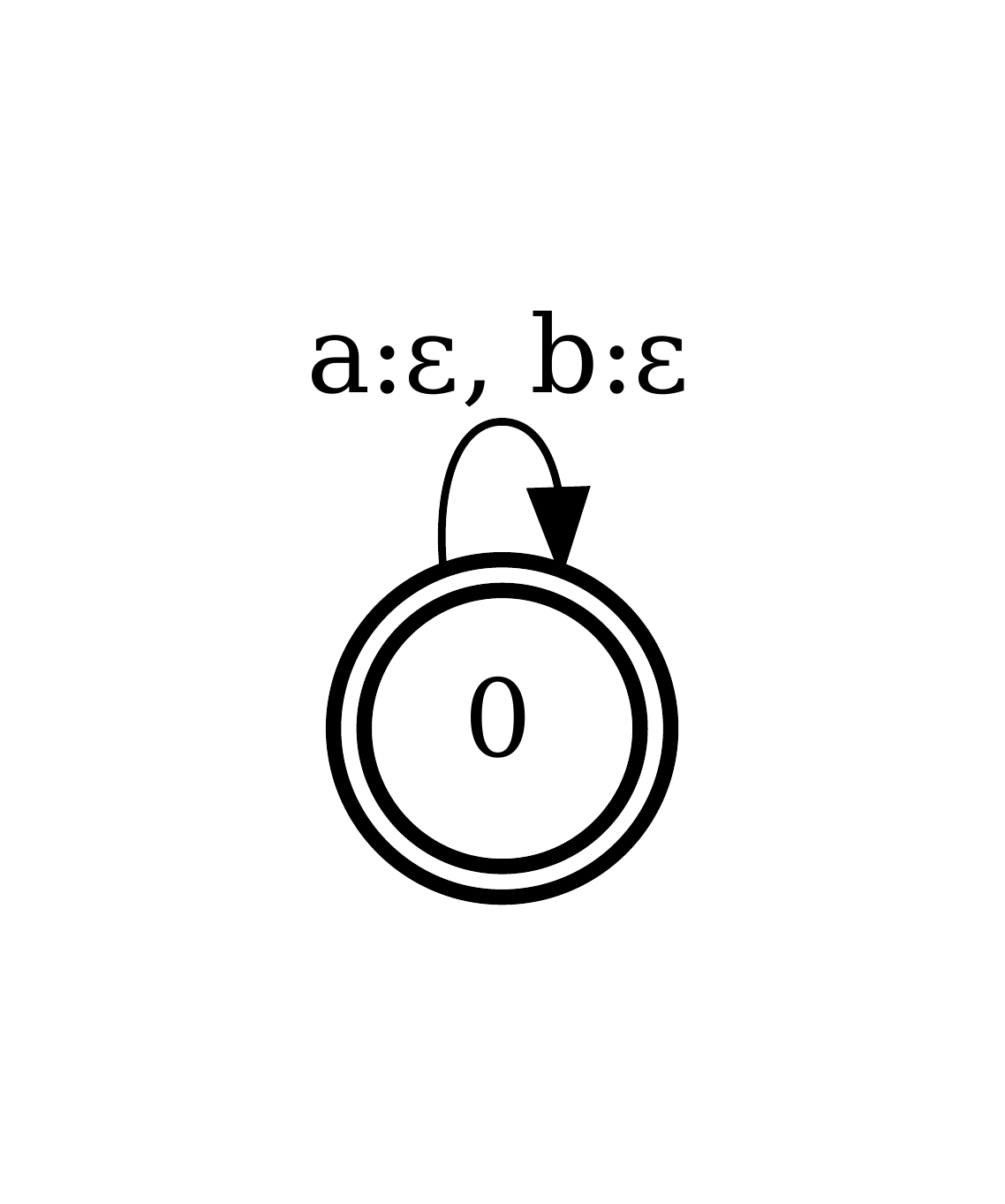} %
                \caption{The eraser transducer. (\texttt{eraser}) }
                \label{fig:A2}
            \end{subfigure}\\
      \begin{subfigure}[b]{\columnwidth}
        \centering
        \includegraphics[width=\textwidth]{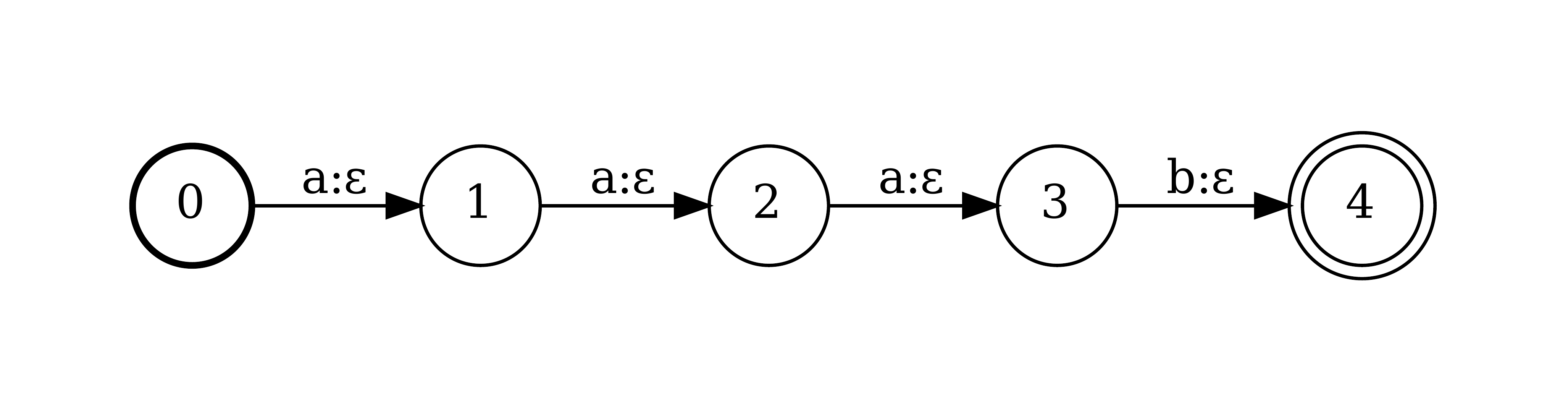}
        \caption{The acceptor for \texttt{aaab} after composition with the eraser. (\texttt{pattern\_erased})}
        \label{fig:B}
      \end{subfigure}
    \end{tabular}
    &
    \begin{subfigure}[b]{\columnwidth}
        \hspace*{-0.7cm}
      \includegraphics[scale=0.09]{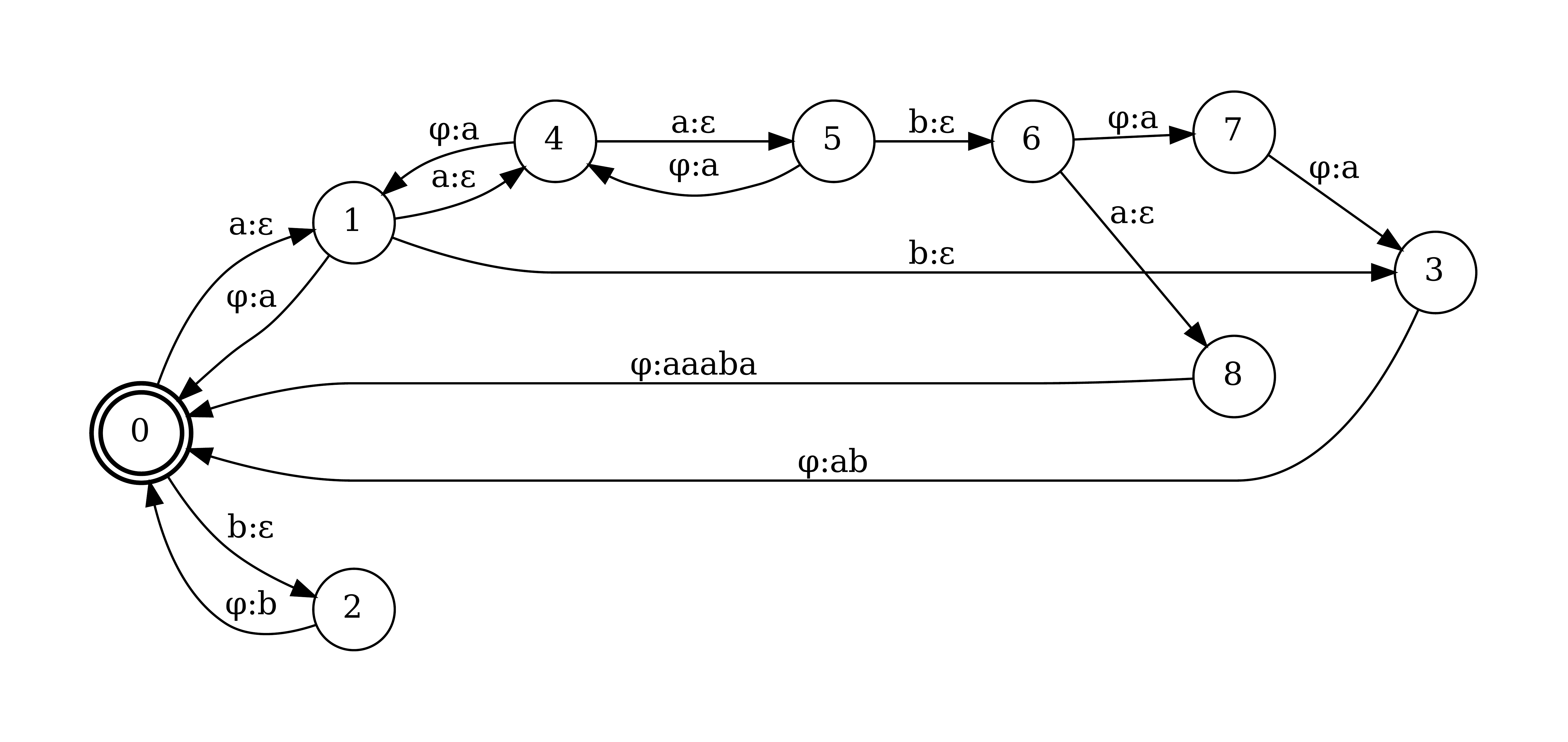}
      \vspace{0.2cm}
      \caption{The Aho-Corasick Trie for $\Gamma = \{\texttt{a}, \texttt{b}, \texttt{ab}, \texttt{aaaba}\}.$ \hspace{\textwidth} (\texttt{transducer})}
      \label{fig:C}
    \end{subfigure}\\
        \begin{tabular}[b]{c}

            \begin{subfigure}[b]{\columnwidth}

                \hspace*{-0.4cm}\includegraphics[scale=0.1]{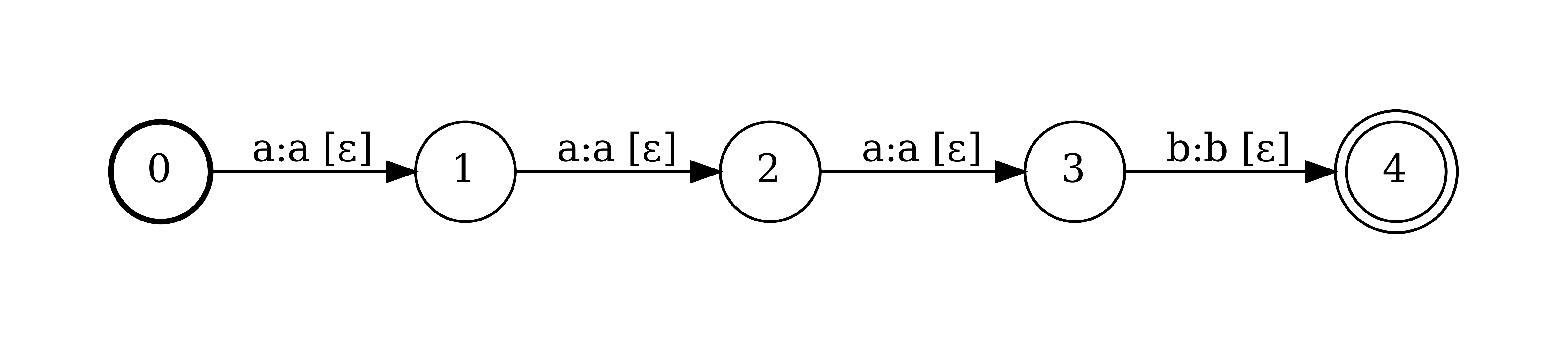}
                \vspace*{0.4cm}                
                \caption{The transducer from (c) mapped to the Gallic semiring. (\texttt{pattern\_gal})}
                \label{fig:linear_gal}
            \end{subfigure}
        \end{tabular}
        &
        \begin{tabular}[b]{c}
            \begin{subfigure}[b]{\columnwidth}
                \hspace*{-0.8cm}
                \includegraphics[scale=0.09]{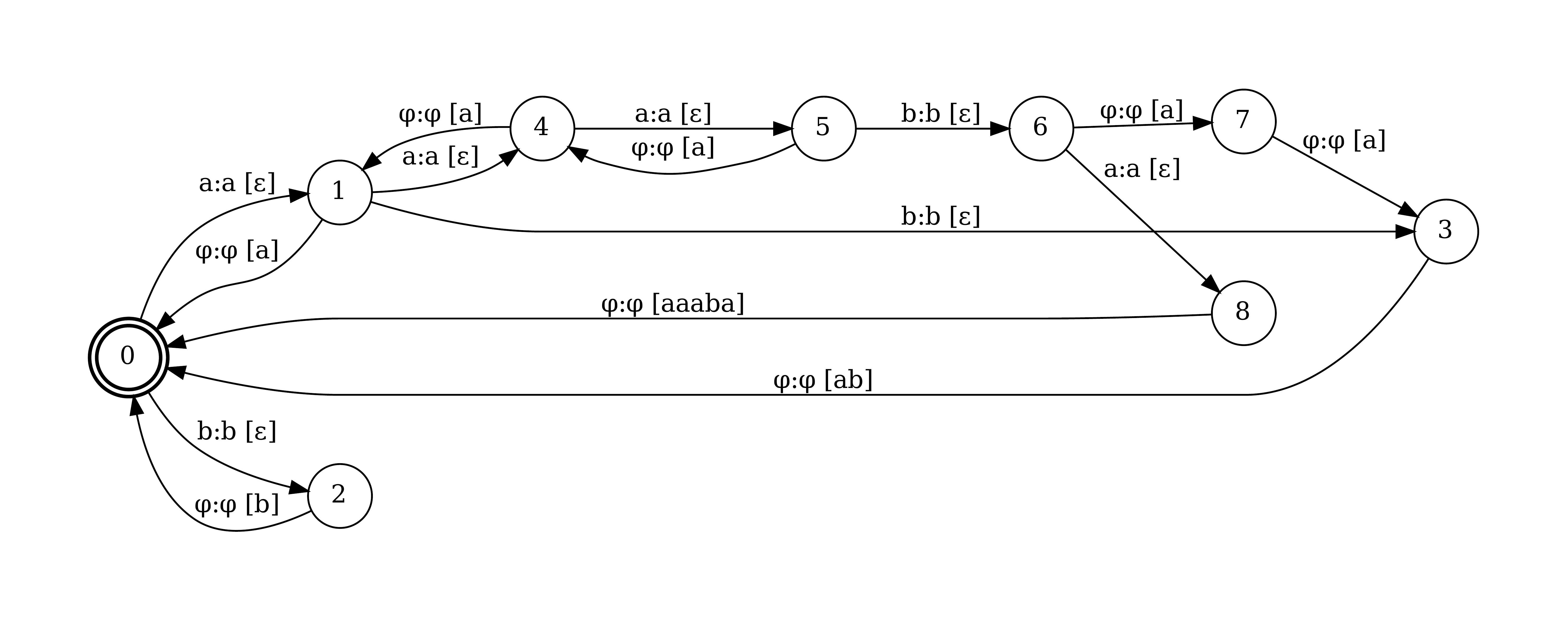}
                \caption{The Aho-Corasick Trie mapped to the Gallic semiring. (\texttt{transducer\_gal})}
                \label{fig:aho_trie_gal}
            \end{subfigure}
        \end{tabular}\\ \hline
      \multicolumn{2}{c}{ %
      \begin{subfigure}[c]{\columnwidth}
        \hspace*{-0.4cm}\includegraphics[scale=0.1]{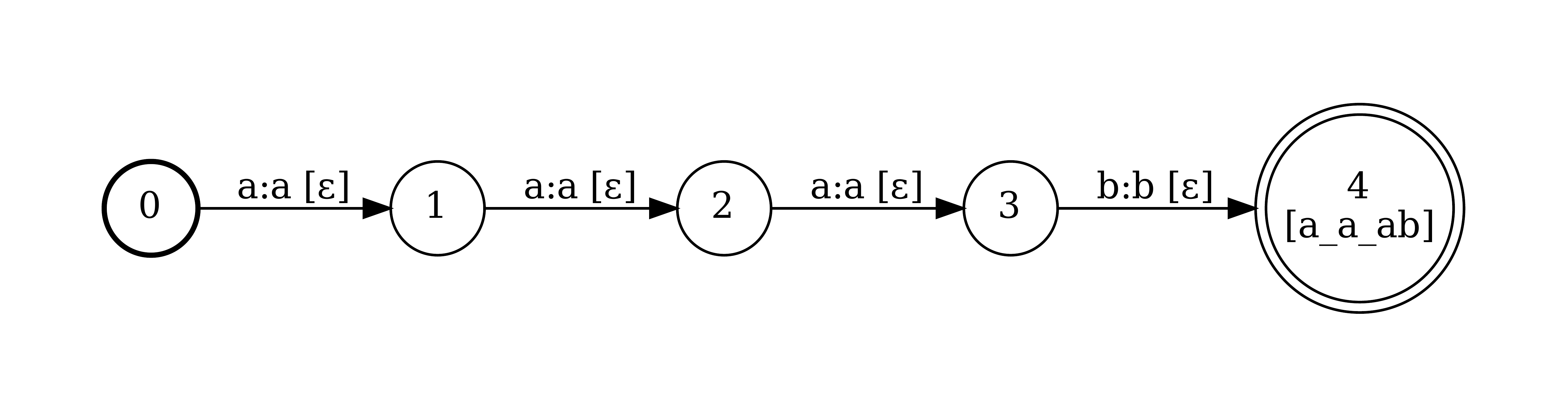}
        \vspace*{-0.8cm}
        \caption{The composition (e) $\circ$ (f) in the Gallic semiring. (\texttt{composed\_gal})}
        \label{fig:A}
      \end{subfigure}
     }\\ \hline
        
     \multicolumn{2}{c}{ %
      \begin{subfigure}[c]{\columnwidth}
        \hspace*{-0.35cm}
        \includegraphics[scale=0.1]{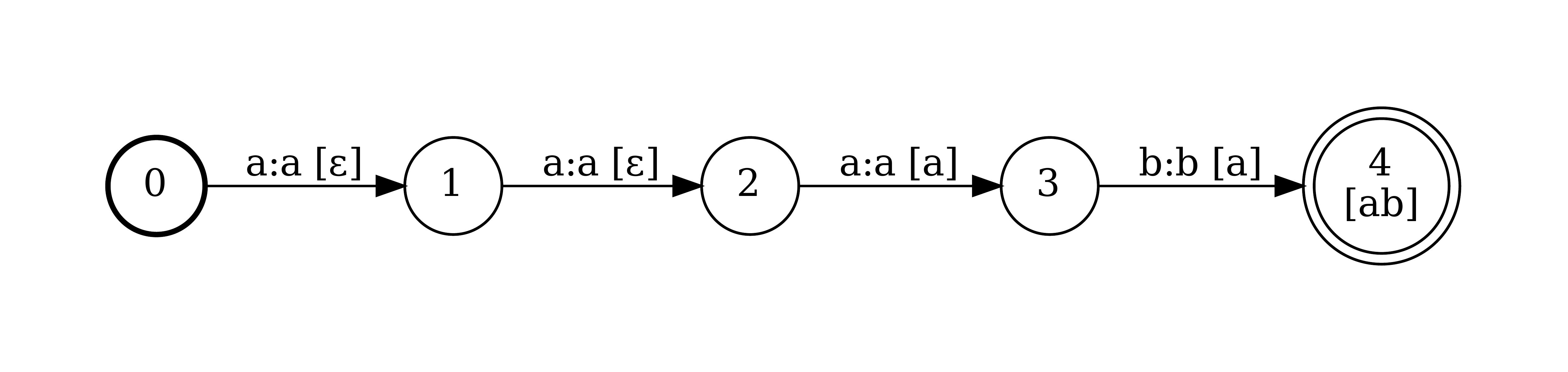}
        \vspace*{-0.8cm}
        \caption{The result of (g) after weight factorization.\hspace{\linewidth} (\texttt{factored})}
        \label{fig:A}
      \end{subfigure}
     }\\ \hline
    \multicolumn{2}{c}{ %
      \begin{subfigure}[c]{\columnwidth}
        \hspace*{0.2cm}
        \includegraphics[scale=0.1]{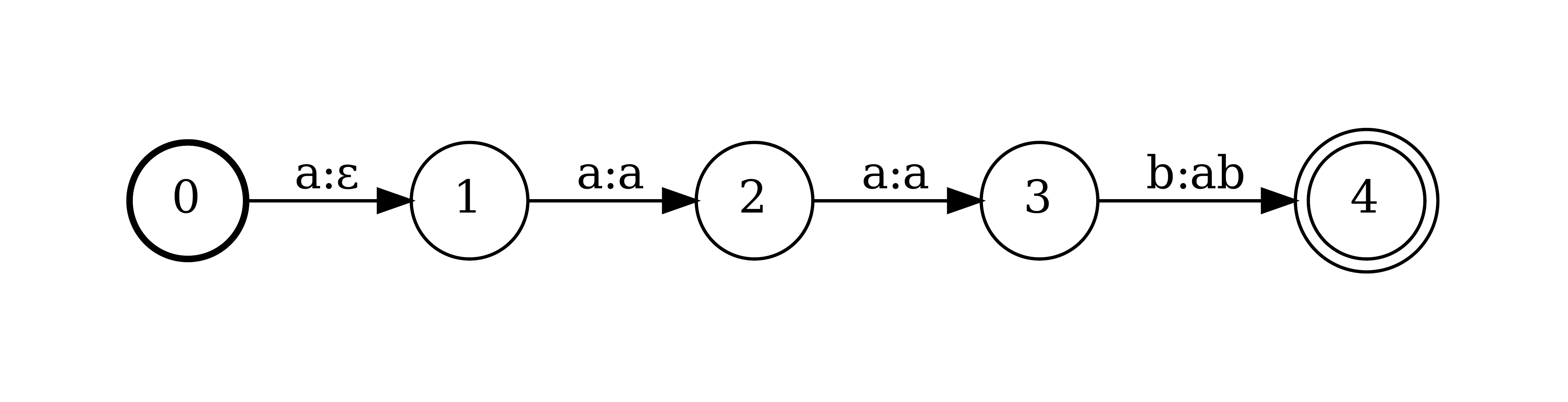}
        \vspace*{-0.5cm}
        \caption{The result of (h), mapped back from the Gallic semiring. (\texttt{converted\_back})}
        \label{fig:A}
      \end{subfigure}
     }\\ \hline
      \multicolumn{2}{c}{ %
      \begin{subfigure}[c]{\columnwidth}

        \hspace*{0.6cm}
        \includegraphics[scale=0.1]{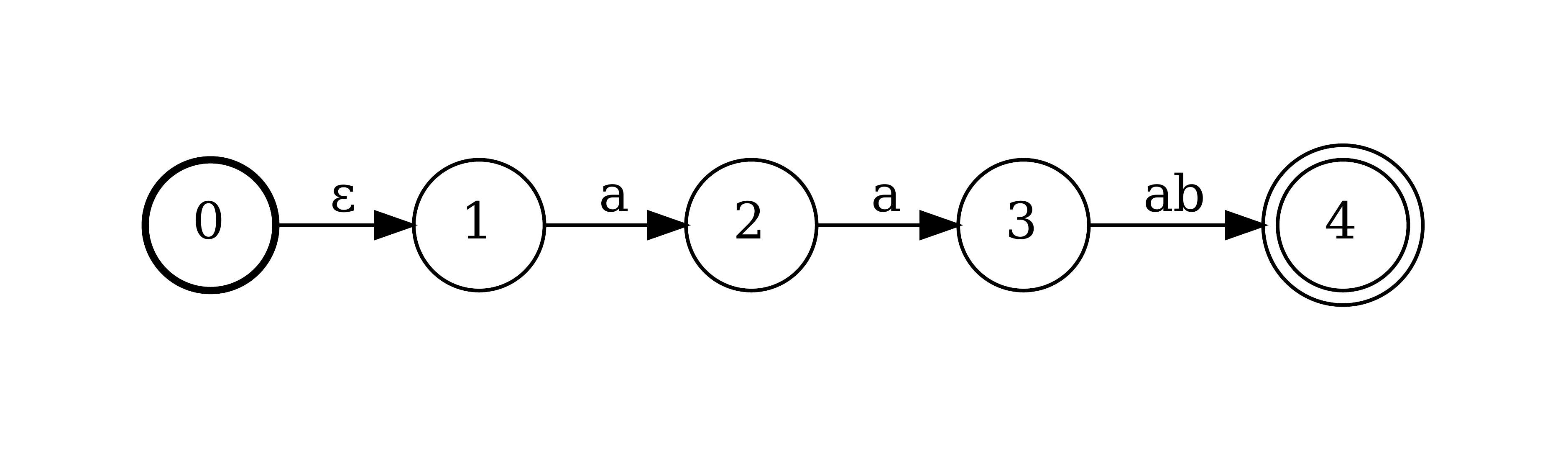}
        \vspace*{-0.5cm}
        \caption{The result of (i), projected to the output labels. (\texttt{composed\_proj})}
        \label{fig:A}
      \end{subfigure}
     }\\ \hline
      \multicolumn{2}{c}{ %
      \begin{subfigure}[c]{\columnwidth}

        \centering
        \hspace*{0.15cm}\includegraphics[scale=0.13]{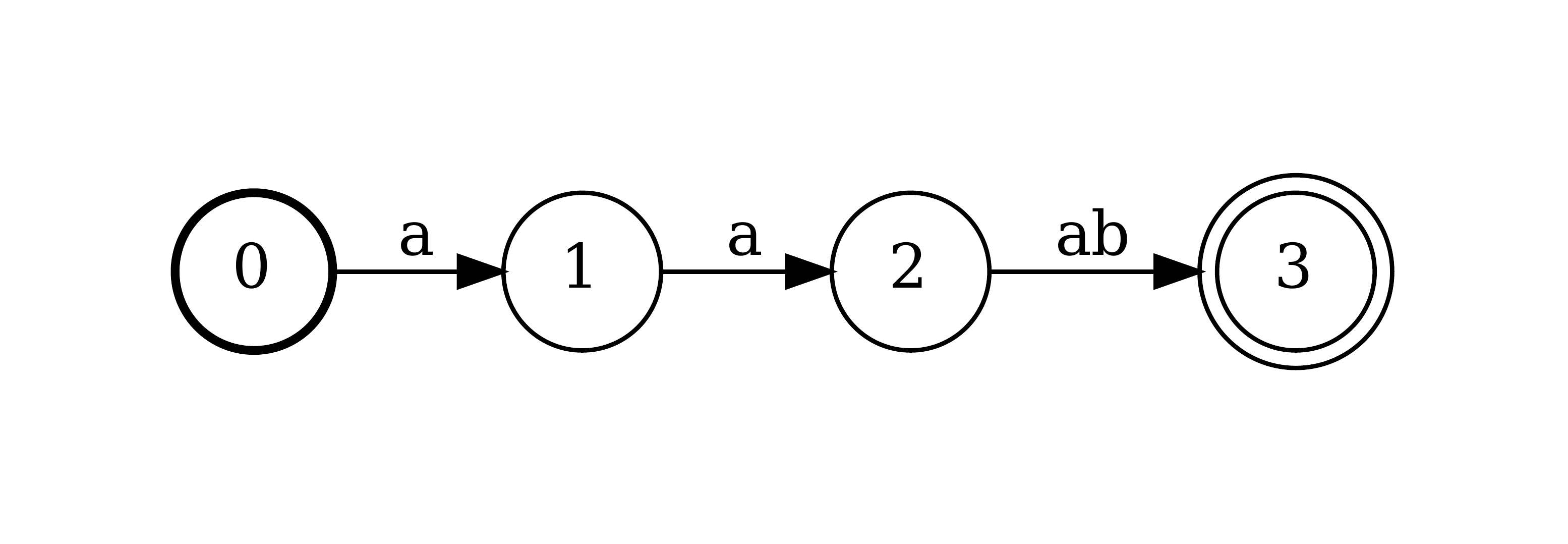}
        \vspace*{-0.5cm}
        \caption{The result of (j) after minimization. (\texttt{composed\_proj\_rm\_eps} and \texttt{det})}
        \label{fig:A}
      \end{subfigure}
     }
     
  \end{tabular}
  \caption{A walk-through of each step of the $\varphi$-Transduction from Code Snippet \ref{code:correct_implementation}. Each subfigure contains a description of the transducer and the corresponding variable name from Code Snippet \ref{code:correct_implementation}. Input and output symbols are denoted as \texttt{i:o} (and the presence of just one symbol implies the form \texttt{i:i}). Weights (only for the string weight part of the Gallic semiring transducers) are placed in [square brackets]. All omitted weights are assumed to be 1, and states with double-circle have weight 1 (for the Gallic semiring transducers, the double circle signifies the non-string Gallic weight being 1, and the Gallic string weight is listed in brackets if it is not $\varepsilon$).}\label{fig:walkthrough}
\end{figure*}

\begin{Code Snippet*}[!p]
       \thisfloatpagestyle{empty}
    \vspace{-1cm}
    \begin{minted}[fontsize=\small]{c++}
    using namespace fst;
    typedef ArcMapFst<StdArc, GallicArc<StdArc>, ToGallicMapper<StdArc>> AMF;
    typedef PhiMatcher<Matcher<AMF>> PM;
    StdVectorFst PhiCompose(const StdVectorFst& pattern, 
                            const StdVectorFst& transducer, 
                            const SymbolTable& syms) {

          // A transducer that maps v -> <epsilon> for all v in the symbol table.
          // We omit the construction for brevity, but it is a single-state
          // FST with an arc (0, v, <epsilon>, 0) for all v in the symbol table.
          auto eraser = OutputEraser(syms);
          
          // The input pattern but with outputs erased. This is to allow the output labels
          // to commute in the Gallic semiring (see `impl_opts.allow_noncommute` below).
          auto pattern_erased = ComposeFst<StdArc>(pattern, eraser);
        
          // Convert the input pattern (after output erasure) and transducer to the 
          // Gallic semiring (both of these are lazy operations).
          ArcMapFst pattern_gal(pattern_erased, ToGallicMapper<StdArc>());
          ArcMapFst transducer_gal(transducer, ToGallicMapper<StdArc>());
        
          // Setting up the composition options (the matchers, etc.).
          ComposeFstImplOptions<Matcher<AMF>, PM> impl_opts;
        
          // A generic matcher for the left (pattern) side of the composition.
          impl_opts.matcher1 = new Matcher<AMF>(pattern_gal, MATCH_NONE);
          
          // A phi matcher for the Gallic semiring on the right side (transducer) 
          // of the composition.
          impl_opts.matcher2 = new PM(transducer_gal, MATCH_INPUT, syms.Find("<phi>"));
          
          // Recall that the string semiring (and therefore the Gallic semiring) does
          // not commute, so we have to manually disable the non-commutative check.
          // But, the zero of the string semiring commutes with every element so disabling
          // this check is ok.
          impl_opts.allow_noncommute = true;
        
          /* The main composition code. The following operations are all lazy (on-the-fly). */
          
          // Compose the pattern and transducer in the Gallic semiring.
          auto composed_gal = ComposeFst<GallicArc<StdArc>>(pattern_gal, transducer_gal, impl_opts);
              
          // Recover the factored weights (in this case, the sequence of 
          // labels in the Gallic semiring string weight).
          auto factored = FactorWeightFst<GallicArc<StdArc>, 
                                          GallicFactor<typename StdArc::Label,
                                          StdArc::Weight>>(composed_gal);
              
          // Revert the factored transducer back to the standard semiring from the Gallic
          // semiring (i.e., move the string weight components back to the arc labels).
          auto converted_back = ArcMapFst(factored, FromGallicMapper<StdArc>());
          
          // Project to the output labels (which hold the output labels from the transducer).
          auto composed_proj = ProjectFst<StdArc>(converted_back, ProjectType::OUTPUT);
          
          /* Now the output is an automaton (with <epsilons>) that corresponds to the
             maximal-match token sequence of the input. */
    
          // Remove <epsilons>, this is a lazy operation.
          auto composed_proj_rm_eps = RmEpsilonFst<StdArc>(composed_proj);
          
          // Determinize the result (this is not a lazy operation).
          StdVectorFst det;
          Determinize(composed_proj_rm_eps, &det);
          
          // Minimize the result (this is not lazy and is done in-place [destructively]).
          Minimize(&det);
          
          return det;
        }
    \end{minted}
    \caption{A complete implementation of $\varphi$-Transductions via the Gallic semiring. See Figure \ref{fig:walkthrough} for a visual example.}\label{code:correct_implementation}
\end{Code Snippet*}

\afterpage{\addtocounter{page}{-1}}

\subsubsection{Commutativity in the Gallic Semiring}
Transducer composition is typically required to be over a commutative semiring \cite{DBLP:conf/wia/AllauzenM08}, but the string semiring, and by extension the Gallic semiring, is not commutative over $\otimes$ (string concatenation). However, in any semiring, the zero element commutes with all other elements. For the string semiring, the one element is the empty string, and $1_{\mathbb{S}} \otimes w = \varepsilon w = w \otimes 1_{\mathbb{S}} = w\varepsilon = w$. Thus, any Gallic weight of the form $(1_{\mathbb{S}}, x_{\mathbb{T}})$ commutes as long as $\mathbb{T}$ is a commutative semiring.

\subsection{Arc Mappers\footnote{\url{https://www.openfst.org/twiki/bin/view/FST/ArcMapDoc}}\textsuperscript{,}\footnote{\url{https://www.openfst.org/twiki/bin/view/FST/FstAdvancedUsage\#ArcMappers}}}

In \textsc{OpenFst}, \textit{arc mappers} are used to apply a transformation on arcs \underline{and final weights} (despite the name) in a transducer. For example, they can be used to perform a modification of all weights in a transducer or to map to a new semiring. Both lazy (\texttt{ArcMapFst}) and non-lazy (\texttt{ArcMap}) arc map implementations are available.

Arc mappers are predicated on a mapping function, several of which are provided by \textsc{OpenFst}. In particular, the \texttt{ToGallicMapper} and \texttt{FromGallicMapper} arc mappers, which transform an arc (over semiring $\mathbb{T}$) to/from the Gallic semiring by extracting the string representation of the output label of the arc to form the string semiring weight, and vic4e versa.

\subsection{Factor Weights\footnote{\url{https://openfst.org/doxygen/fst/html/factor-weight_8h_source.html}}}
Factor weights are designed to transform a single weight $w$ into a sequence of weights $w_1 \otimes w_2 \otimes \dots \otimes w_n = w$ which will be placed on successive arcs. In particular, we use it with the Gallic semiring to factor a string weight representing a sequence of symbols into a sequence of arcs, each with one symbol on it. For example, if the Gallic weight of an arc had string weight \texttt{ab\_a}, it would be factored into two arcs with weights \texttt{ab} and \texttt{a}, respectively.

\section{The \textsc{OpenFst} Implementation}

We now introduce a way to use the Gallic semiring to properly implement $\varphi$-transductions in \textsc{OpenFst}. The general idea is to exploit the fact that \texttt{PhiMatchers} correctly accumulate \textit{weights} along $\varphi$-paths (one or more in a chain). So, by treating the \textit{symbols} as weights (which is what the Gallic semiring does), we can accumulate symbols across $\varphi$-transitions and then map the weights back to labels (using factor weights), giving the desired result. \textsc{OpenFst} provides functionality for all of this, and all of the operations we need have a \textit{lazy} implementation: they do not fully materialize the transducers, but instead construct them on-the-fly. This allows us to efficiently chain together our transformations and compute the final result.

A complete implementation with notes is given in Code Snippet \ref{code:correct_implementation}.

\subsection{MaxMatch $\varphi$-Transductions}
    
We demonstrate an end-to-end implementation of $\varphi$-transductions with the Gallic Semiring by implementing the MaxMatch (WordPiece) tokenization algorithm that is commonly used in natural language processing \cite{devlin-etal-2019-bert,song-etal-2021-fast}. Given a vocabulary $\Gamma$ made up of words\footnote{Typically, we use \textit{subwords}, which are word parts, or word pieces (hence the name of the algorithm). For example, a subword vocabulary might contain \texttt{bike} and \texttt{s}, which can be combined to encode the string \texttt{bikes} as $\texttt{[bike][s]}$.} from $\Sigma^*$ (where $\Sigma \subseteq \Gamma$), we want to convert sequences from $\Sigma^*$ into sequences from $\Gamma^*$, in a \textit{greedy} fashion --- at each stage, we pick the longest word in $\Gamma$ which matches the text and use that as the next token.

A connection between the Aho-Corasick algorithm, MaxMatch tokenization, and finite-state automata was described (but not implemented) by \newcite{song-etal-2021-fast}, where they show that the a modified Aho-Corasick pattern matching automaton \cite{Aho:72} can be used to implement MaxMatch tokenization in linear time. The key insight is that the Aho-Corasick automaton contains \textit{failure} arcs, which can be traversed whenever no suitable transition exists at a state in order to go to a new state which corresponds to the longest proper suffix of the current string that is a prefix of a word in the vocabulary. Additionally, each state includes a \textit{popping} information, which encodes the longest prefix of the currently unmatched string which corresponds to a token in the vocabulary. When processing a string, one walks through the automaton until reaching a state where no suitable transition exists. The popping information tells us which elements of $\Gamma$ to output and the failure arc tells us which state to transition to in order to continue the traversal without having to backtrack. Combining the popping and failure arcs as $\varphi$-transitions, where a failure transition is labeled $\varphi\texttt{:w}$ if $\texttt{w}$ is the popping information word and the target state is the state described by the failure function, gives a  MaxMatch $\varphi$-transducer.

The MaxMatch $\varphi$-transducer can be used to tokenize strings by first encoding the string as a finite-state automaton and then performing composition. It can also be used to tokenize \textit{languages} (i.e., patterns described by an automaton over $\Sigma$) in the same way. The result is language over $\Gamma$ where each subword sequence matches the input pattern (after converting back to the character-level representation), and corresponds to the greedy, MaxMatch tokenization.

\bibliography{anthology,custom}
\bibliographystyle{acl_natbib}

\appendix
\begin{Code Snippet*}[htbp]
    \begin{minted}[fontsize=\small]{c++}
    using namespace fst;
    StdVectorFst PhiAutomatonCompose() {
      typedef PhiMatcher<Matcher<StdVectorFst>> PM;
      SymbolTable syms;
      syms.AddSymbol("<epsilon>");
      syms.AddSymbol("<phi>");
      syms.AddSymbol("a"); syms.AddSymbol("b"); syms.AddSymbol("c");
      
      StdVectorFst left;
      left.AddStates(2);
      left.SetStart(0);
      left.SetFinal(1);
      left.AddArc(0, StdArc(syms.Find("a"), syms.Find("a"), 1));
      
      StdVectorFst right;
      right.AddStates(3);
      right.SetStart(0);
      right.SetFinal(2);
      right.AddArc(0, StdArc(syms.Find("<phi>"), syms.Find("<phi>"), 1));
      right.AddArc(1, StdArc(syms.Find("a"), syms.Find("a"), 2));

      ComposeFstImplOptions<Matcher<StdVectorFst>, PM> impl_opts;
      impl_opts.matcher1 = new Matcher<StdVectorFst>(left, MATCH_NONE);
      impl_opts.matcher2 = new PM(right, MATCH_INPUT, syms.Find("<phi>"));
      
      auto composed = ComposeFst<StdArc>(left, right, impl_opts);
      StdVectorFst det;
      Determinize(composed, &det);
      Minimize(&det);
    
      return det;
    }
    \end{minted}
    \caption{An implementation of $\varphi$-transitions when composing \textit{automata}. This corresponds to Figures \ref{fig:sub1}-\ref{fig:sub3}, and correctly outputs the language $\{\texttt{[a]}\}$.}\label{code:phi_automata}
\end{Code Snippet*}

\begin{Code Snippet*}[htbp]
    \begin{minted}[fontsize=\small]{c++}
    using namespace fst;
    StdVectorFst IncorrectPhiTransducerCompose() {
      typedef PhiMatcher<Matcher<StdVectorFst>> PM;
      SymbolTable syms;
      syms.AddSymbol("<epsilon>");
      syms.AddSymbol("<phi>");
      syms.AddSymbol("a"); syms.AddSymbol("b"); syms.AddSymbol("c");
      
      StdVectorFst left;
      left.AddStates(2);
      left.SetStart(0);
      left.SetFinal(1);
      left.AddArc(0, StdArc(syms.Find("a"), syms.Find("a"), 1));
      
      StdVectorFst right;
      right.AddStates(3);
      right.SetStart(0);
      right.SetFinal(2);
      right.AddArc(0, StdArc(syms.Find("<phi>"), syms.Find("c"), 1));
      right.AddArc(1, StdArc(syms.Find("a"), syms.Find("b"), 2));

      ComposeFstImplOptions<Matcher<StdVectorFst>, PM> impl_opts;
      impl_opts.matcher1 = new Matcher<StdVectorFst>(left, MATCH_NONE);
      impl_opts.matcher2 = new PM(right, MATCH_INPUT, syms.Find("<phi>"));
      
      auto composed = ComposeFst<StdArc>(left, right, impl_opts);
      StdVectorFst det;
      Determinize(composed, &det);
      Minimize(&det);
    
      return det;
    }
    \end{minted}
    \caption{{\color{red} NB: This code does not work correctly and is only included for reference.} The same implementation as Code Snippet \ref{code:phi_automata}, but for \textit{transducer} composition with $\varphi$-transitions. Despite being intuitive and similar to the automata case, it is incorrect. The expected result is a transducer that encodes $\texttt{[c][b]}$ as the output, but the actual result is the language $\{\texttt{[b]}\}$.}\label{code:incorrect}
\end{Code Snippet*}

\section{The Incorrect Method}\label{apx:why_doesnt_it_work}

Code Snippet \ref{code:incorrect} gives a minimal (not) working example of trying to transduce using regular \texttt{PhiMatchers} and composition in \textsc{OpenFst}. It creates two transducers, one which accepts the string \texttt{a} and one which should transduce $\texttt{[a]} \rightarrow \texttt{[c][b]}$. However, the actual result is the language $\{\texttt{[b]}\}$ (after projection to the output labels).

Other variants of this (for example, by adding a $\varphi$ loop at each state of the left-side automaton in order to match the $\varphi$ on the right-side transducer) generally fail for similar reasons: \texttt{PhiMatcher} ignores the output label on $\varphi$-arcs and do not work well with $\varphi$-arcs followed by $\varepsilon$-arcs. This due to two design decisions in \textsc{OpenFst}: 1) since $\varepsilon$ is non-consuming, there is always a match for it at any state and 2) $\texttt{PhiMatchers}$ only accumulate \textit{weights} and not \textit{labels} when traversing $\varphi$-transitions. On the other hand, \texttt{PhiMatchers} cannot accumulate labels because it would require outputting multiple labels on a match which is not possible within the \texttt{Matcher} API. This is the inspiration for converting to the Gallic semiring and then accumulating the string weights across $\varphi$-transitions.

\end{document}